\documentclass[journal]{IEEEtran}
\pdfoutput=1
\usepackage{cite}

\usepackage{color}

\ifCLASSINFOpdf
  \usepackage[pdftex]{graphicx}
  \graphicspath{{../pdf/}{../jpeg/}}
  \DeclareGraphicsExtensions{.pdf,.jpeg,.png}
\else
  \usepackage[dvips]{graphicx}
  \graphicspath{{../eps/}}
  \DeclareGraphicsExtensions{.eps}
\fi
\usepackage{epstopdf}
\usepackage{subfigure}

\usepackage{amsmath}
\newtheorem{Def}{Definition}
\makeatletter

\newcommand{\Rmnum}[1]{\expandafter\@slowromancap\romannumeral #1@}
\makeatother

\usepackage{algorithmic}
\usepackage[linesnumbered, ruled, vlined]{algorithm2e} 

\begin{document}
%
\title{A Region-based Collaborative Management Scheme for Dynamic Clustering in Green VANET}
%
%

\author{Bingyi Liu, Zhipeng Fang, 
Wei Wang,
Xun Shao,
Wei Wei,
Dongyao Jia,
Enshu Wang,
Shengwu Xiong 		
		\thanks{Manuscript received XXX, XX, XXXX; revised XXX, XX, XXXX. This work was supported by National Natural Science Foundation of China (No. 61802288) and Major Technological Innovation Projects in Hubei Province, China (No. 2019AAA024).}
		
		\thanks{Bingyi Liu, Zhipeng Fang
		and Shengwu Xiong are with the School of Computer Science and Technology, Wuhan University of Technology, Wuhan, China.}
        \thanks{Wei Wang is with the School of Intelligent Systems Engineering, Sun Yat-Sen University, Shenzhen 510275, China. (email: ehomewang@ieee.org)}
        \thanks{Xun Shao is with the School of Engineering, Kitami Institute of Technology, Kitami 0908507, Japan. (email: x-shao@ieee.org)}
        \thanks{Wei Wei is with School of Computer Science and Engineering, Xi'an University of Technology, Xi'an 710048, China (email: weiwei@xaut.edu.cn).}
        \thanks{Enshu Wang is with the University at Buffalo, The State University of New York, Buffalo, NY, USA.}
        \thanks{Dongyao Jia is with the School of Civil Engineering, The University of Queensland, Brisbane, Australia.}
	}

%
%

\markboth{Journal of \LaTeX\ Class Files}%
{Shell \MakeLowercase{\textit{et al.}}: Bare Demo of IEEEtran.cls for IEEE Journals}
%



\maketitle

\begin{abstract}
Green Vehicular Ad-hoc Network (VANET) is a newly-emerged research area which focuses on reducing harmful impacts of vehicular communication equipments on the natural environment. Recent studies have shown that grouping vehicles into clusters for green communications in VANETs can significantly improve networking efficiency and reduce infrastructure costs. As a dynamic network system, maintaining the network connectivity and reducing the communication overlap are two critical challenges for green VANET clustering. However, most existing work studies connectivity and overlap separately, lacking a deep understanding of the relationship between them. To address this issue, we present a comprehensive analysis that jointly considers the two critical factors in one model. Specifically, we first design a state resemblance prediction (SRP) model based on the historical trajectory feature relevance between vehicles; Combined with the SRP model, we propose the region-based collaborative management scheme (RCMS) to establish the dynamic clustering; Lastly, we take extensive experiments to verify the region-based collaborative management scheme for dynamic clustering. The results demonstrate that the proposed clustering algorithm can achieve high networking efficiency and better communication stability.
\end{abstract}


%
\IEEEpeerreviewmaketitle

\section{Introduction}
%
%
%
%

\IEEEPARstart{V}{A}NETs enable direct communication among vehicles in a vehicle-to-vehicle (V2V) manner and promote to form a green and swarm intelligent network system in which vehicles perform as network nodes. With the help of V2V communication, vehicles on the roads can share different types of information (e.g., traffic situations, road accidents) to improve road safety and traffic efficiency, allowing for a more environmentally friendly smart and sustainable transportation system\cite{lyu2019characterizing}.
 

Similar to the communications in internet of things (IoTs) \cite{chen2021compressed,li2019ls,yu2021blockchain,yu2021secure}, when facing the continuously increasing traffic demand, the fundamental challenge in greening communication and networks then turns to be: can we and how can Send More Information bits with Less Energy (SMILE) \cite{niu2020green}. Meanwhile, the high dynamic and density of vehicles in urban traffic environments often lead to frequent communication disconnection and high communication overhead. Inspired by cloud computing and fog computing\cite{li2018virtual,yang2020distributed}, some existing studies utilize a hierarchical network structure by grouping vehicles into clusters instead of a flat network structure\cite{liu2019hierarchical}. Within the cluster, ordinary vehicles communicate directly with core vehicles, and the core vehicles implement inter-cluster communication.

Obviously, for such a hierarchical network structure, it is critical to design an effective dynamic clustering algorithm. 
Some traditional clustering algorithms, e.g., beacon-based clustering \cite{nguyen2016efficient, naderi2019adaptive}, mobility-based clustering \cite{2019Mobility, ren2017mobility}, and backbone-based clustering \cite{ucar2015multihop}, are designed based on large-scale message dissemination and need to update the clustering states at a fixed period continually by utilizes k-hop cluster architecture to enhance clustering efficiency. Moreover, some prediction-based clustering algorithms \cite{benzerbadj2018cross,lalitha2017gccr,ullah2019advances,lin2016mozo,jaiswal2017location} and learning-based networking schemes \cite{chen2021self,chen2021adversarial} have been proposed based on nodes' spatial-temporal properties, which help to form the relevance among the current position and potential destinations, and determines the cluster or networking structure.
Although the aforementioned studies are fundamentally vital for clustering algorithms, several issues have not been fully addressed. \textcolor{black}{First, most existed clustering algorithms are devoted to implementing the clustering construction session but slightly consider the cluster maintenance aspect. Since the highly dynamic network topology severely affects the original network structure and the uneven distribution of vehicles gives rise to high communication overhead, cluster maintenance is considered as an indispensable procedure of clustering in VANETs. Besides, few studies consider the clustering method with an in-depth understanding of the relationship between network connectivity\cite{lyu2018dbcc} and communication overlap. For network connectivity, most existing studies focus on optimization based on the mobility of vehicles, e.g., relative positions and relative speed, to maintain network connectivity. However, complex urban traffic may severely affect vehicle mobility and network topology, resulting in excessive communication overlap and network congestion.}




\textcolor{black}{In this paper, we propose a region-based collaborative management scheme for dynamic clustering, named RCMS, aiming at improving the clustering performance in terms of maintaining the network connectivity and reducing the communication overlap. Such a task gives challenges from the following aspects. \textcolor{black}{Firstly, the high dynamics of vehicles in real-world traffic environments may cause the instability of network topology, resulting in low network quality and high communication overhead. Thus, It is critical to adjust the network structure as the environment evolves. Secondly, the management of network structure often requires additional computation and communication resources. The limited onboard resources pose challenges to the communication process of the whole system so that the resources can be fully utilized.}
In RCMS, we clarify how the metrics of network connectivity and communication overlap affect network quality and communication overhead. Also, We jointly consider the two critical factors into one model, intending to achieve low end-to-end latency and stable communications.
The main contributions of this paper are summarized as follows.}
\begin{itemize}
    \item We propose the state resemblance prediction (SRP) model for obtaining the trajectory features relevance between vehicles. According to the SRP model, we can group the vehicles with similar trajectory features to improve the clustering performance.
    \item Based on the SRP model, we propose a region-based collaborative management scheme for dynamic clustering in VANET. Specifically, we propose the region-based collaborative management scheme to jointly consider the relationship between the network connectivity and communication overlap. According to the analysis, we provide both intra-region and inter-region procedures and aim at achieving better network quality and lower communication overhead.
    \item We simulate real traffic and communication scenarios on a simulation platform and take extensive experiments to evaluate networking efficiency and communication performance. The results validate the efficiency of the region-based collaborative management scheme for dynamic clustering.
\end{itemize}

The rest of this paper is organized as follows. Section \Rmnum{2} summarizes the related work, and Section \Rmnum{3} presents the system model and problem description. We then present a state resemblance prediction method in Section \Rmnum{4} and analyze the performance of the region-based collaborative management scheme in Section \Rmnum{5}, respectively. Finally, the paper is concluded in section \Rmnum{6}.

\section{RELATED WORK}

Using vehicle clustering to support green vehicular communications is an important technique that focuses on reducing overhead and costs of VANETs. Recently, various clustering algorithms have been extensively proposed in the literature, e.g. \cite{cooper2016comparative,hasrouny2017vanet}. Considering our work is highly relevant to the clustering algorithm, in this section, we first introduce traditional clustering algorithms and then, in particular, focus on predictive clustering algorithms.

\subsection{Traditional Clustering Algorithms}

Traditional clustering algorithms normally update the clustering states continually via large-scale message dissemination\cite{wang2018networking,liu2017infrastructure}, and utilize k-hop cluster architecture to enhance clustering efficiency. The various traditional clustering algorithms are classified as follows.

As for typical beacon-based clustering algorithms, clusters are constructed based on the beacon messages of network parameters detected by receiving vehicle, containing the parameters, e.g., the vehicle speed, vehicle density, and connect time \cite{liu2017joint}. The cluster-based beacon dissemination process (CB-BDP) in \cite{allouche2015cluster} provides vehicles with a local vehicle proximity map of their nearby vehicles, informing drivers of hazard situations to avoid accidents. To cope with the channel contention caused by the fixed beacon interval, \cite{lyu2020towards} has proposed a full distributed adaptive beacon control scheme to avoid the rear-end collision in dense scenarios based on individually estimated danger coefficient.

Density-based clustering can provide robust connectivity within a cluster according to the density information about the composition of the cluster. In \cite{lyu2018momac}, the author proposes a mobility-aware TDMA MAC, named MoMAC, which can allocate each vehicle a time slot according to vehicles’ mobility and road topology. The mobility-based clustering aims to minimize relative mobility among the vehicle and thereby maintain clustering convergence, and dynamics \cite{2019Mobility}. For example, Wang \emph{et al.} \cite{wang2015vanet} takes into account the impact of wireless channel conditions, MAC protocol, and vehicle mobility, and proposes the closed-form expressions about throughput and average packet loss probability of a VANET cluster. Similar to \cite{ren2016new}, \cite{ren2017mobility} chooses the central node in the cluster as the cluster head to improve the stability of cluster management. 

In recent years, some studies focus on k-hop clustering algorithms to improve the clustering efficiency and stability by reducing the variation in cluster head and cluster member lifetime. Each cluster chooses a node as the cluster head, and the distance between cluster head and cluster member can be one or more hops \cite{ucar2015multihop}. For example, \cite{senouci2019mca} allows vehicles to connect to the VANET through a roadside unit so that each vehicle can obtain and share the necessary information about its multi-hop neighbors to perform the clustering process. The clustering stability can be enhanced by electing both slave cluster head and master cluster head. 

\subsection{Predictive Clustering Algorithms}

\begin{figure*}[htb]
\centering
\includegraphics[width=0.9\textwidth]{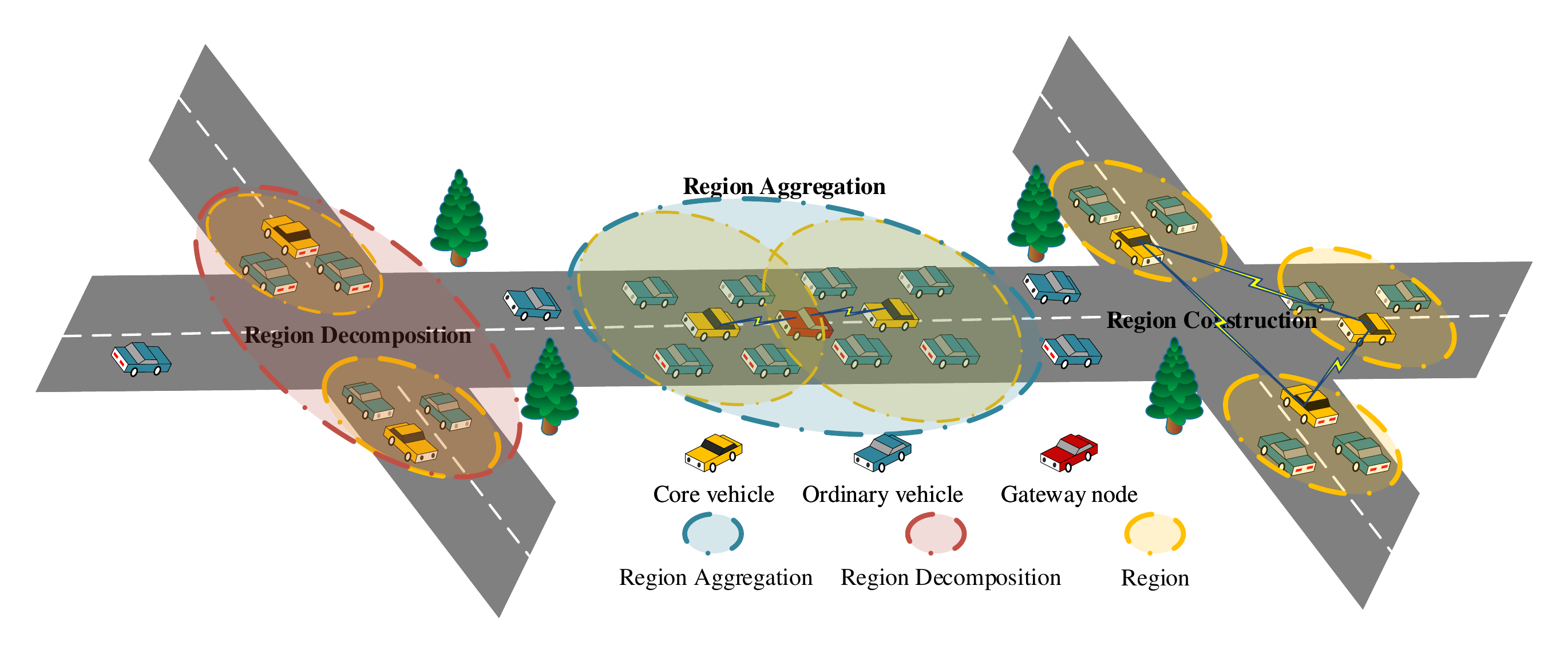}
\caption{Region-based collaborative management scheme for VANET in urban scenarios} 
\label{scenarios}
\end{figure*}

The predictive clustering algorithms predict the movement of a vehicle based on current geographic information and vehicle states. Some of these algorithms can be categorized as follow.

The main idea for position-based clustering algorithms is that vehicles are grouped into a cluster on the basis of geographic position among the vehicles \cite{benzerbadj2018cross, yao2017v2x}. To analyze the network topology with the vehicular position information, Qu \emph{et al.} proposes a dynamically evolving networking (DEN) \cite{qu2020complex} to take realistic vehicular traces as input and explores the impact of node addition, node deletion, and link loss on the network due to node movement. As for the connectivity-based clustering algorithms, due to the spatial-temporal properties of the vehicle, the network connectivity is important to indicate the networking efficiency. In \cite{cheng2020connectivity}, the author proposes an algorithm named connectivity prediction method (CP), which combines the communication range and vehicle density to build up a connectivity prediction model for dynamic clustering.  
For understanding the connectivity probabilities under different communication manners, the author \cite{shao2015performance} analyzes and predicts the connectivity probabilities for the Vehicle-to-Vehicle (V2V) and Vehicle-to-Infrastructure (V2I) communications on one direction and bi-direction road, respectively. Recently, some proposed predictive clustering algorithms have taken advantage of fog computing. The paper \cite{ullah2019advances} presents an FoG-oriented architecture that utilizes road junctions for route discovery and vehicles in the parking area for packet transmission. 

Although the clustering algorithms mentioned above are essential to support clustering, most existing studies only consider network connectivity during the cluster construction phase, regardless of its importance in the cluster maintenance phase. Furthermore, the highly dynamic network topology severely affects the original network structure; dealing with communication area overlap gradually becomes a high priority in the cluster maintenance phase. In addition, most studies treat connectivity and overlap as separate metrics without consideration of their intrinsic connection. This paper proposes a region-based collaborative management scheme for VANETs in urban scenarios. Moreover, the joint control schemes of overlap and connectivity can achieve better clustering management and have wider adaptability especially in complex urban scenarios. 
\section{SYSTEM MODEL AND PROBLEM DESCRIPTION}

In this section, we first describe the traffic scenario and communication model in an urban area and then give a description of the vehicle movement model. To facilitate the following descriptions, we first summarize the notations in Table~\ref{table:1}.

\subsection{Traffic Scenario and Communication Model}

\textcolor{black}{In this study, we consider a typical traffic scenario that consists of roads and intersections. As illustrated in Fig.~\ref{scenarios}, a hierarchical network structure is proposed for dynamic clustering. Such a hierarchical network structure is suitable for large-scale VANETs in urban environments. Compared with the traditional flat network structure, the hierarchical network can achieve a dynamic balance of networking efficiency and communication stability Specifically, considering unique characteristics of vehicular networks, such as highly heterogeneous resources and highly dynamic system status, the hierarchical network structure alleviates the high coupling among vehicles in a flat network structure by grouping vehicles into clusters. Moreover, it enables logically centralized control by decoupling management and communication planes, and enhances system reliability, scalability and flexibility. In this paper, the hierarchical network structure consisted of three layers. The layer of core vehicles manages relationships among regions and needs to conduct region operations including region construction, region aggregation, and region decomposition. The layer of gateway vehicles is responsible for connection among regions. The layer of ordinary vehicles focuses on direct communication and information exchanges in the same region.}

To support the region operation, we assume that all vehicles are equipped with on board units (OBU) and global positioning systems (GPS) for capturing the kinetic dynamic states (e.g., speed, acceleration, position) and realizing V2V communication. The messages are disseminated among the vehicles based on IEEE 802.11p standard. Each vehicle enters VANETs with its own unique identity and meets the requirement of strict security and privacy. 

\subsection{Vehicle Movement Model}
\begin{figure*}[htb]
\centering
\includegraphics[width=0.9\textwidth]{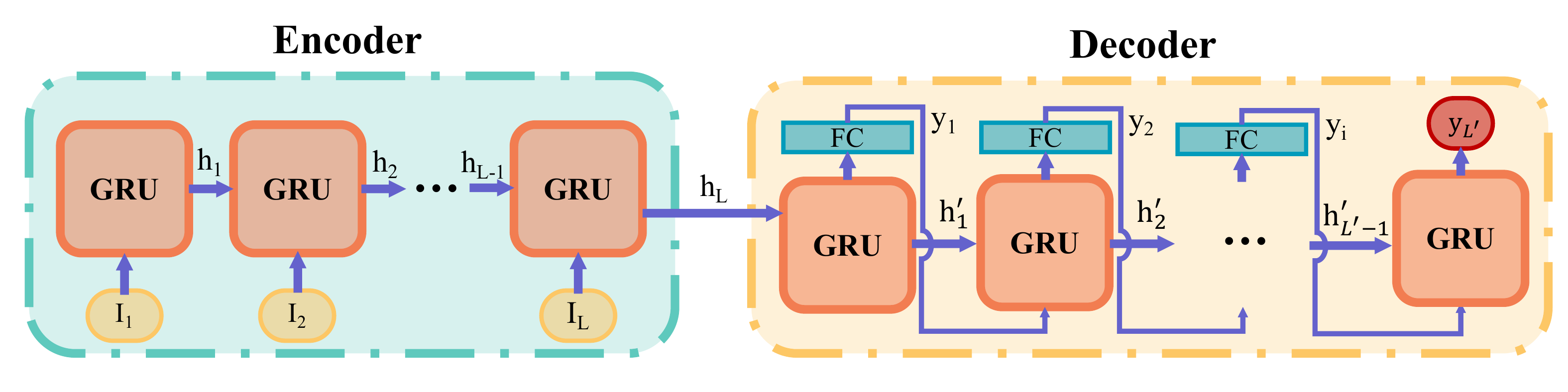}
\caption{the state-resemblance-prediction framework} 
\label{framework}
\end{figure*}

\begin{table}
\textcolor{black}{
\caption{NOTATION}
\centering
\begin{tabular}{p{1.2cm} p{5.8cm}} 
\hline
\hline
$\mathbf{Symbol}$ & $\mathbf{Description}$\\
\hline
$v^\kappa_i$ & Vehicle $i$, $i$ is the ID of vehicle, $\kappa$ denotes the vehicle type.\\
\hline
$s_i(t)$ & the state of vehicle $v^\kappa_i$ at the time $t$\\
\hline
$\gamma_i$ & the road segment $v^\kappa_i$ stays\\
\hline
$\Vec{s_i}$ & the vehicle $v^\kappa_i$'s speed\\
\hline
$\Vec{a_i}$ & the vehicle $v^\kappa_i$'s acceleration\\
\hline
$\delta_{TTI}$ & the traffic condition\\
\hline
$\theta$ & the vehicle's driving direction\\
\hline
$\zeta$ & the estimated interaction time for received message\\
\hline
$\lambda_{SRP}$ & the similarity between vehicle $v^o_i$ and $v^c_i$ trajectories\\
\hline
$h_{t}$ & the hidden state at time $t$\\
\hline
$\tilde{h}_{t}$ & the candidate hidden state at time $t$\\
\hline
$z_{t}$ & the update gate\\
\hline
$r_{t}$ & the reset gate\\
\hline
$\tau_{coo}$ & the cooperative threshold among vehicle $v^o_i$ and $v^c_i$\\
\hline
$\tau_{com}$ & the competitive threshold among vehicle $v^o_i$ and $v^c_i$\\
\hline
$\varphi_{agg}$ & the region aggregation threshold among vehicle $v^c_i$ and gateway node\\
\hline
$\varphi_{dec}$ & the region decomposition threshold among vehicle $v^o_i$ and $v^c_i$\\
\hline
\hline
\end{tabular}
\label{table:1}
\vspace{.2cm}
}
\end{table}
Our objective is to propose a region-based collaborative management scheme that integrates multiple vehicles with similar historical and current states. Each core vehicle is responsible for region operations and message dissemination. In the following content, we formulate vehicles' states and vehicle trajectory information.
\begin{Def} \label{Def:state}
the states of vehicle node $v^\kappa_i$ at time $t$ denoted as
\begin{align}
    {state_i}(t) = \{\kappa, \gamma_i, \Vec{s_i}, \Vec{a_i}, \theta, \delta_{TTI} \}\nonumber
\end{align}
\end{Def}
For the vehicle node $v^\kappa_i$, $i$ means the unique identity, and $\kappa$ denotes the vehicle type. Other notations about the vehicle's features are described as follows, $\gamma_i$ denotes the road segment that the vehicle stays at the time $t$, ${\Vec{s_i}}$ represents the vehicle's speed, $\Vec{a_i}$ denotes the acceleration of the vehicle. In addition, $\theta$ represents the vehicle's driving direction, which is a relative value referring to the core vehicle's driving direction. If the vehicle is driving in the same direction as the core vehicle, the value of $\theta = 1$. Otherwise, $\theta = -1$. In order to quantify traffic conditions, we introduce the concept of the travel time index (TTI), which calculated by in Equation~(\ref{Eq:TTI}),
\begin{equation}\label{Eq:TTI}
\delta_{TTI} = \frac{\Sigma_{i=1}^{N_r} \frac{\gamma_{i}}{\Vec{s_i}} \cdot \omega_{i}}{\sum_{i=1}^{N_r} \frac{\gamma_{i}}{V_{\text {free}_{i}}} \cdot \omega_{i}}
\end{equation}
where $\omega_{i}$ denotes the weight of the road segment $\gamma_i$ based on different urban road hierarchy, and $N_r$ represents the number of road segments. $V_{\text {free}_{i}}$ is the free flow of $\gamma_i$. The $\delta_{TTI}$ is the evaluation index, i.e., the ratio of the actual travel time and the free flow time, and can reflect the urban congestion degree. A larger value of the index implies a worse traffic congestion level. 
\begin{Def} \label{Def:trajectory}
The trajectory information of each vehicle is defined as
\begin{align}
    T_{v^\kappa_{i}}(n) = \{(state_i(t_0), t_0), (state_i(t_1), t_1),\nonumber \\ \nonumber \dots, (state_i(t_n), t_n)\}\nonumber
 \end{align}
\end{Def}
where $T_{v^\kappa_i}$ denotes vehicle $v^\kappa_i$ trajectory, and $state_i(t_n)$ represents vehicle $v^\kappa_{i}$ states at time $t_n$.

\section{STATE RESEMBLANCE PREDICTION MODEL}

In this section, we first overview the proposed SRP model and then detail overall training algorithm of the SRP model.

\subsection{Framework Overview}

Since vehicles' trajectory features are time-dependent, we formulate the prediction process of vehicles' trajectory features as the time series forecasting tasks. Due to the ability of automatic feature extraction, the Recurrent Neural Network (RNN) is very effective in processing data with time-series features. Meanwhile, we use a smooth approximation based on dynamic time warping (DTW)\cite{cuturi2017soft} to train our RNN. Compared with the traditional loss function, like MSE, DTW can better capture the difference between the two trajectories. Hence, we integrate the Sequence To Sequence (Seq2Seq) RNNs model with the DTW algorithm in the SRP model. As illustrated in Fig.~\ref{framework}, the proposed framework utilizes two GRU networks, namely the encoder-decoder framework, consisting of the encoder module and decoder module. 

\subsubsection{GRU module}
GRU is a highly effective variant of RNN, which alleviates the sequence-based tasks with long-term dependencies in RNN. To alleviate these adverse effects, GRU applies the update and reset gates. The activation $h_{t}$ restricts updating process between the previous activation $h_{t-1}$ and the current candidate activation $\tilde{h}_{t}$ is shown in Equation~(\ref{Eq:ht}),
\begin{equation}\label{Eq:ht}
    h_{t}=z_{t} \odot h_{t-1}+\left(1-z_{t}\right) \odot \tilde{h}_{t}
\end{equation}
where the update gate $z_t$ decides how much the previous moment $h_{t-1}$ is updated to the current $h_{t}$ given in the following Equation~(\ref{Eq:zt}),
\begin{equation}\label{Eq:zt}
    z_{t}=\sigma\left(W_{z} x_{t}+U_{z} h_{t-1}+b_{z}\right)
\end{equation}
where $W$, $U$, and $b$ are the parameters of our model, the different subscripts correspond to different formulas. Moreover, their values are updated during training. The matrix $W$ multiplies the vectors by $x_t$ and produces a new vector. Similarly, the matrix $U$ multiplies the vectors by $h_{t-1}$ and produces a new vector. These two new vectors and the parameter bias are summed, and each component of the result is passed to the sigmoid function. Further, Equation~(\ref{Eq:ht-1}) mainly calculates the element-wise multiplication of the reset gate $r_t$ and $h_{t-1}$, which determines the previous information being retained or forgotten,
\begin{gather}
\tilde{h}_{t}=\tanh \left(W_{h} x_{t}+U_{h}\left(r_{t} \odot h_{t-1}\right)+b_{h}\right)\label{Eq:ht-1}\\
r_{t}=\sigma\left(W_{r} x_{t}+U_{r} h_{t-1}+b_{r}\right)\label{Eq:rt}
\end{gather}
The reset gate $r_t$ primarily determines how much past information needs to be forgotten, as given in Equation~(\ref{Eq:rt}).

\subsubsection{Encoder module}
The encoder module obtains the surrounding ordinary vehicles' moving status to build the trajectory sample and obtains its vehicle status information to generate the vector information. The Encoder module encodes the input sequence $\{I_1, I_2, \dots, I_L\}$ of length $L$ of trajectory state information into the corresponding hidden states and obtains the final output $h_{L-1}$ after $L$ times recursive loops. Then, the encoder module transfers $h_L$ to the decoder module, and the decoder module uses it as the initial cell state for the following sequence generation. 

\subsubsection{Decoder module}
The Decoder uses vector information from the encoder module to detect the trajectory's state resemblance and recursively generates subsequent trajectory samples. The Decoder Module commences with its previous output as the Decoder input $h_{L}$. Besides, each updating process necessitates taking the last moment output $h^{\prime}_{l-1}$ as the input and connects the fully connected network through the activation function called Rectified Linear Unit (ReLU). Furthermore, the decoder module recursively generates the output sequence $\{O_1, O_2, \dots, O_n\}$. 

\subsection{Overall Training Algorithm}

In this part, we present the overall training algorithm of SRP model. For an ordinary vehicle $v^o_i$, we use the DTW to compare the two time-series dissimilarities given in the following Equation~(\ref{Eq:DTW}),
\begin{equation} \label{Eq:DTW}
    DTW(T_{v^c_i}, T_{v^o_i})=\min \frac{\sqrt{\sum_{i=1}^{n} p_i}}{n}
\end{equation}
where $p_i$ is defined as the warping path of length $n$. The $i\,th$ element of $p$ is defined as $p_k = (i, j)$, which corresponds to the mapping of trajectory sequence. $T_{v^c_i}$ and $T_{v^o_i}$ denote the trajectory of core vehicles and ordinary vehicles, respectively. To handle the trajectory information, we investigate the vehicle trajectories continuously and capture vehicle nodes' movement information among road segments from the temporal and spatial perspective. 


Due to the non-differentiable property of DTW, inspired by the perspective proposed in \cite{Guen2019shape}, we define the objective function combining DTW as shown in Equation~(\ref{Eq:object function}),
\begin{equation}\label{Eq:object function}
    \xi_{SRP}(T_{v^c_i}, T_{v^o_i}) = -\log \left(\sum_{i=1}^{L}  \exp \left(-\frac{DTW(T_{v^c_i}, T_{v^o_i})}{L^2}\right)\right)
\end{equation}
\textcolor{black}{where $L$ denotes the length of trajectory $T_{v^\kappa_i}$. For the trajectory information of length $L$, the vehicle dynamics characteristics are included in the form of time series, effectively alleviating the vehicle dynamics challenges by combining the forward and backward transfer in our model. During the training process, our objective function compares the ordinary vehicle trajectory $T_{v_i^o}$ with the core vehicle trajectory $T_{v_i^c}$. The trajectory distortion terms are based on the alignment between the ordinary vehicle trajectory $T_{v_i^o}$ and the core vehicle trajectory $T_{v_i^c}$. Our loss function, based on DTW,  focuses on the dissimilarity between core and ordinary vehicles' trajectory information by temporally aligning the trajectory series between core and ordinary vehicles. In DTW, the warping path is based on a binary matrix $M \subset \{0,1\}^{L^2}$, where $M_{c,o} = 1$ when the trajectory $T_{v^c_i}$ is associated with $T_{v^o_i}$, otherwise $M_{c,o} = 0$.}

\section{THE REGION-BASED COLLABORATIVE MANAGEMENT SCHEME}
In this section, we provide a complete region-based collaborative operations scheme that monitors metrics, e.g., communication overlap and network connectivity, and performs appropriate region management to guarantee clustering performance, including (1) region construction, (2) core vehicle replacement, (3) region maintenance.

\subsection{Region Construction}
Region construction is triggered when a vehicle connects to the VANETs. The vehicles will select a suitable region nearby or form their own region. The region construction criteria are based on the similarity of the mobility metrics and trajectory features. In the following, we present the main procedures of region construction in detail.
\SetKwInOut{CoreVehicle}{Core Vehicle $v^c_i$}
\SetKwInOut{OrdinaryVehicle}{Ordinary Vehicle $v^o_i$}
\begin{algorithm}[t]
\caption{Region Construction}
\label{alg:region constuction}
\SetAlgoLined
\OrdinaryVehicle\
    $v^o_i$ send READ message to one-hop vehicles \; 
    $v^o_i$ start WaitingTime() // waiting for the return message \;
\CoreVehicle\
    $v^c_i$ receive the READ message from ordinary vehicle \;
\eIf{$\delta_{TTI} > 1.5$}
    {$v^c_i$ send ROGER messages to two-way vehicles \;}
    {$v^c_i$ send ROGER messages to same direction vehicles \;}
\OrdinaryVehicle\  
\eIf{WaitingTime() $< \zeta $ and $v^o_i$ Receive the ROGER message from core vehicle }{
    \uIf{one ROGER message is received}
        {$v^o_i$ send cooperative request to core vehicle \;}
    \uElseIf{more than one ROGER message is received}{
        \For{Each core vehicle in contact with the $v^o_i$}{
            $v^o_i$ calculate the cooperative threshold ${\tau}_{coo}$ using Equation~(\ref{Eq:taucoo}) \;
        }
        $v^o_i$ send cooperative request to core vehicle with the highest cooperative threshold \;
    }
}{
    $v^o_i$ reset the WaitingTime() and enter the waiting mode \;
}
\CoreVehicle\
    $v^c_i$ receive the cooperative request from ordinary vehicle \;
    $v^c_i$ send a agreement to cooperative ordinary vehicle \;

\end{algorithm}

\textcolor{black}{First, an ordinary vehicle sends a message, named $READ$, to its surrounding one-hop vehicles, triggering a timer for calculating the waiting time. Next, the ordinary vehicle waits for the return message $ROGER$ from the core vehicle within the time $\zeta$, which contains the core vehicle's unique identifier. The $\zeta$ denotes the estimated interaction time for the received message. When receiving the core vehicle message, the ordinary vehicle will respond individually based on the number of $ROGER$ messages: 
\begin{itemize}
\item If the number of $ROGER$ packets is 0, the vehicle enters the waiting mode and waits for the next $\zeta$ interval to send the $READ$ message again.
\item If the number of $ROGER$ packets is 1, the vehicle directly regards the packet's sender as its core vehicle.
\item If the number of $ROGER$ packets is greater than one, the ordinary vehicle will select the core vehicle with the highest cooperative threshold to join according to the Equation~(\ref{Eq:taucoo}).
\end{itemize}}

It shall be noted that, when receiving the $ROGER$ message from the multiple core vehicles, the ordinary vehicle will evaluate the cooperative threshold, respectively. In this situation, the ordinary vehicle has the ability to communicate with multiple core vehicles, and we define this kind of ordinary vehicle as a gateway node. 
\textcolor{black}{The cooperative threshold between vehicle $v^o_i$ and corresponding $v^c_i$ is calculated by Equation~(\ref{Eq:taucoo}),
\begin{equation}\label{Eq:taucoo}
{\tau}_{coo} =\left\{
\begin{aligned}
\exp (-\Delta{\Vec{s_{ij}}} * \Delta{d_{ij}}) & , & 1<\delta_{TTI}<=1.5, \\
\lambda_{SRP} * \exp (-\Delta{\Vec{s_{ij}}} * \Delta{d_{ij}}) & , & \delta_{TTI}>1.5.
\end{aligned}
\right.
\end{equation}
The above equation takes into account the influence of three factors: the relative speed ${\Delta{\Vec{s_{ij}}}}$, the relative distance $\Delta{d_{ij}}$, and the similarity of vehicle trajectories $\lambda_{SRP}$. A higher similarity means that the ordinary vehicle may stay longer with the core vehicle. Further, the basic idea of $\delta_{TTI}$ is the ratio of free-flow speed and actual speed. 
According to \cite{kong2015measuring}, based on the value of TTI, the traffic conditions can be classified into six different degrees (e.g., traffic tie-up, congestion, light congestion, slow, smooth, free flow). In this paper, the TTI value above 1.5 indicates congestion or even traffic jam. Otherwise, it means that the traffic is smooth or free-flow conditions. Specifically, the value of $\delta_{TTI}$ can be set in two ranges:
\begin{itemize}
    \item If $\delta_{TTI}$ is less than 1.5, it means that the traffic is smooth in this situation. To mitigate the problem of link disconnections due to the fast movement of vehicles, we evaluate the influence of relative speed and distance on the cooperative threshold. Ordinary vehicles that maintain lower relative mobility metrics to core vehicles are more suitable for maintaining the network's stability.
    \item If $\delta_{TTI}$ is greater than 1.5, it indicates that the overall traffic flow is slow. In this condition, the core vehicle can adequately handle the trajectory similarity $\lambda_{SRP}$ with ordinary vehicles to select more suitable ordinary vehicles for the region and alleviate the communication overhead. 
\end{itemize}}

After calculating the cooperative threshold, the ordinary vehicle selects the core vehicles with the highest score, sending the cooperative request. Then, the core vehicle will return an agreement to the ordinary vehicle. The complete region construction procedure is described in Algorithm~\ref{alg:region constuction}.

\subsection{Core Vehicle Replacement}
\SetKwInOut{CoreVehicle}{Core Vehicle $v^c_i$}
\SetKwInOut{OrdinaryVehicle}{Ordinary Vehicle $v^o_i$}
\begin{algorithm}[t]
\caption{Core Vehicle Replacement}
\label{alg: Replacement}
\SetAlgoLined
\CoreVehicle\
    \While{$v^c_i$ send message to corresponding ordinary vehicles}{
        $v^c_i$ check the OVNum() //number of the ordinary vehicles \;
    }
\OrdinaryVehicle\
    $v^o_i$ receive the message from core vehicle \;
    \uIf{$\delta_{TTI} < 1.5$}{
       $v^o_i$ update the new information every updating interval \;
    }
    \uElseIf{$\delta_{TTI} > 1.5$}{
        \eIf{Change in velocity of more than $50\%$ within the current timestamp}{
            $v^o_i$ update newly information every updating interval \;
        }{
            $v^o_i$ update newly information in two consecutive updating interval \;
        }
    }
\CoreVehicle\  
\If{more than half of reduction of ordinary vehicles in the region}{
    \uIf{more than one candidate core vehicle}{
        \For{Each candidate core vehicle}{
            $v^c_i$ calculate the competitive threshold ${\tau}_{com}$ using Equation~(\ref{Eq:taucom}) \;
        }
        $v^c_i$ send request to ordinary vehicle with the highest threshold \;
    }
    \uElseIf{a suitable candidate core vehicle is not available}{
        \textbf{goto Region Constuction}
    }
}
\end{algorithm}
As time passed, the high dynamic of the vehicle may impact the region's stability. The choices of core vehicles are probably no longer optimal. Hence, in this situation, we will find a new vehicle to replace the previous core vehicle for maintaining the intra-region structure's stability.

In the region, the core vehicle will adjust communication policy with the ordinary vehicle according to different $\delta_{TTI}$ environments as shown in Algorithm~\ref{alg: Replacement}. When more than half of the ordinary vehicles disconnect with the core vehicle, it means that the current core vehicle deviates from its managed region, it will commence the replacement process. The candidate core vehicle is supposed to maintain a relatively center position within the region and to move at an approximate speed to the ordinary vehicles. Accordingly, we propose the competitive threshold of core vehicles. The competitive threshold includes two dimensions: historical and current driving status data, and can be calculated by Equation (\ref{Eq:taucom}),
\begin{equation}\label{Eq:taucom}
    {\tau}_{com} = \lambda_{SRP} * \frac{\sum_{i=1}^{N_o} {\Delta d_{ij}}}{N_o}
\end{equation}
$\Delta d_{ij}$ represents the distance between the candidate core vehicle and another ordinary vehicle. $N_o$ denotes the number of ordinary vehicles in the same region. The core vehicle has the following two different situations:
\textcolor{black}{
\begin{itemize}
    \item If there is more than one candidate core vehicle, the core vehicle will calculate competitive threshold, respectively. The vehicle with the highest competitive threshold, which has a relatively center intra-region position than the former core vehicle, replaces the former core vehicle.
    \item When the core vehicle fails to communicate with the region due to a significant state change (e.g., the core vehicle makes a U-turn while ordinary vehicles keep the original direction), the core vehicle has no replaceable core vehicle. Hence, the vehicle will select a suitable region nearby or repeats the region construction as shown in Algorithm~\ref{alg:region constuction}, and the original region also selects a new core vehicle.
\end{itemize}}

\subsection{Region Maintenance}


\textcolor{black}{In this paper, network connectivity and communication overlap are two essential metrics for the whole system. The network connectivity can be measured by data transmission rate and delay. While the communication overlap can be measured by the number of vehicles that are applied to connect different clusters. Apparently, The network structure with high overlap means that the connectivity can be enhanced with more vehicles acting as gateway vehicles to connect different clusters. However, the communication overhead may be too high due to the overfrequent exchange of information among clusters. Conversely, the network structure with low overlap may decrease network connectivity and even cause communication disconnection. To improve the networking efficiency and communication overhead, we propose region aggregation and decomposition schemes.}

\subsubsection{Region aggregation}
\SetKwInOut{GatewayNode}{Gateway Node }
\SetKwInOut{NewlyCoreVehicle}{Newly Core Vehicle}
\begin{algorithm}[t]
\caption{Region Aggregation}
\label{alg: Aggregation}
\SetAlgoLined
\GatewayNode\
    Gateway node detect the overlap of the region \;
    \If{more than half of the gateway nodes are in the overlapping region  during two consecutive $\zeta$ periods}{
        \textbf{met the criteria for aggregation operation} \;
        calculate the $\varphi_{agg}$ using Equation~(\ref{Eq:varphi}) \;
        Gateway node send the merging right to vehicle with the highest score \;
    }
\NewlyCoreVehicle\  
    Newly core vehicle broadcast the change of the region, and the former core vehicles become the candidate core vehicle \;

\end{algorithm}
\textcolor{black}{In some traffic situations, the overlap between regions may gradually increase with the dynamic of traffics, leading to the dramatic increase of communication overhead among regions and affect region management efficiency. For the region aggregation operation, the purpose is to alleviate the excessive communication overhead caused by communication overlap, so the different clusters can be integrated into one region.}

The gateway node initiates the region aggregation operation. This is because: (1) As the regions' overlap gradually increases, the gateway node detects this situation's occurrence, indicating region aggregation conditions. (2) The gateway node already obtains information about the core vehicles during the region construction and maintenance sessions. No additional communication overhead is required. 
(3) In the region construction and maintenance session, the gateway node sojourns within the communication range of multiple core vehicles and receives status information from the core vehicles. Specifically, the vehicles within the communication range of multiple core vehicles can be selected as gateway nodes, namely candidate gateway nodes.
In between two consecutive $\zeta$ intervals, if the overlap rate is more than $50\%$, the gateway node will initiate a region aggregation operation. The region aggregation threshold given in Equation (\ref{Eq:varphi}),
\begin{equation}\label{Eq:varphi}
    \varphi_{agg} = \frac{\sum_{i=1}^{N_c} e^{-\eta_i} \cdot \tau_{coo_i}}{\sum_{i=1}^{N_c} {\tau}_{coo_i}}
\end{equation}
where $N_c$ represents the number of core vehicles within the communication range. $\eta_{i}$ indicates the distance between the gateway node and their corresponding core vehicles. The region aggregation operation is given in the following.
\textcolor{black}{
\begin{itemize}
    \item The gateway node detects the overlap of the core vehicle's communication region.
    \item If more than half of the gateway nodes are in the overlapping communication region of the core vehicle during two consecutive $\zeta$ periods, and the distance between the core vehicles does not exceed half of the communication region, the gateway will initiate the aggregation operation. Moreover, the adoption of this strategy can, to a certain extent, alleviate the sudden change in overlap caused by the traffic flow convergence or communication disconnection.
    \item The region aggregation threshold $\varphi_{agg}$ is calculated by the Equation (\ref{Eq:varphi}). The gateway node with the highest $\varphi_{agg}$ takes over the merged region management.
    \item The newly core vehicle broadcasts the change of the region, and the former core vehicles becomes the candidate core vehicle.
\end{itemize}}
The region aggregation process is described in Algorithm~\ref{alg: Aggregation}.

\subsubsection{Region decomposition}

\textcolor{black}{Some ordinary vehicles within the same region tend to leave the communication range due to the different velocities and moving directions at intersections, which may lead to connection instability. For the region decomposition operation, the purpose is to avoid the uneven distribution of vehicles in a region to affect the network connectivity, so the region can be decomposed into several appropriate clusters.}

Here we propose decomposition value in Equation (\ref{Eq:taudec}),
\begin{equation} \label{Eq:taudec}
    \varphi_{dec} = \frac{1-e^{-{\vert \Delta d_{ij} \vert}}}{1+e^{-\vert \Delta({v^c_i}-{v^o_j}) \vert}}
\end{equation}
where $\vert \Delta({v^c_i}-{v^o_j}) \vert$ denotes the relative speed between ordinary vehicles and core vehicles. As the difference in movement patterns between vehicles gradually increases, the distance $\vert \Delta d_{ij} \vert$ between core vehicle and ordinary vehicle also increases, which will affect network connectivity and the region stability. The region aggregation process is described in Algorithm~\ref{alg: decomposition}.

\SetKwInOut{CoreVehicle}{Core Vehicle $v^c_i$}
\SetKwInOut{OrdinaryVehicle}{Ordinary Vehicle $v^o_i$}
\begin{algorithm}[t]
\caption{Region Decomposition}
\label{alg: decomposition}
\SetAlgoLined
\OrdinaryVehicle\
    \If{$v^o_i$ approach an intersection}{
        \textbf{met the criteria for decomposition operation} \;
        \eIf{$v^o_i$ different from the subsequent direction of the moving of the core vehicle}{
            $v^o_i$ leave the corresponding region \;
            \textbf{goto Region Construction} \;
        }{
            $v^o_i$ calculate the $\varphi_{dec}$ using Equation~(\ref{Eq:taudec}) \;
            $v^o_i$ with $\varphi_{dec} < 0$ leave the region \;
        }
    }
\CoreVehicle\  
    $v^c_i$ broadcast the change of the region \;

\end{algorithm}
\section{Experimental Result}

The section evaluates our solution via extensive simulations, we first present the experimental setting, followed by the baseline and parameter settings, and finally, we evaluate the performance of the proposed region-based collaborative management scheme in terms of networking and communication performance. To facilitate further discussions, we summarize the simulation parameters in Table~\ref{table:2}.

\subsection{Experimental Settings}
\begin{table}[t]
\caption{Simulation Parameters}
\centering
\begin{tabular}{p{4.5cm} p{2.5cm}} 
\hline
\hline
$\mathbf{Parameter}$ & $\mathbf{Value}$ \\
\hline
Simulation Time & 650s \\
\hline
Road Topology & 3km$\times$3km \\
\hline
Speed Range & 10 - 30 m/s \\
\hline
Physical/Mac protocol & IEEE 802.11p \\
\hline
Path loss model & Free-space ($\alpha$=2) \\
\hline
Fading model & Nakagami-m (m=3) \\
\hline
Transmission power & 20 dBm \\
\hline
Transmission range & 250m \\
\hline
Date rate & 6Mb/s \\
\hline
Updating Interval & 1s \\
\hline
Waiting Time $\zeta$ & 2s \\
\hline
\hline
\end{tabular}
\label{table:2}
\vspace{.2cm}
\end{table}
In our experiments, we utilize the Veins simulator, which combines OMNeT++ for network simulation and SUMO for the vehicular mobility simulation. For the traffic scenario, unless noted otherwise, we consider a 3km$\times$3km grid in which the road segment is the two-way four-lane with a length of 500m. We set 1200 free-running vehicles in the experiment, and each vehicle has randomly distributed on the road network. Since the experiments simulate an urban scenario, we set the vehicle's maximum speed as 30m/s. We use the Veins and SUMO to simulate scenarios with traffic light control; The vehicles' behavior in the simulator is modeled by a typical car-following model and lane-changing model, and follows the traffic regulations, e.g., obeying the intersection traffic lights. We implement IEEE 802.11p standard for V2V communication, and the maximum communication range is set to 250m.  The region process starts at the 50s until all vehicles entered the map. After the 50s, the vehicle executes the region procedure, and the total simulation time is 650s.

The trajectory data are collected from DiDi Express and DiDi Premier drivers. The measurement interval of the track points is approximately 2-4 seconds. The track points are bound to physical roads to match the trajectory data and the actual road information. (Data source: DiDi Chuxing GAIA Open Dataset Initiative)

\subsection{Dynamic Clustering Simulation Results}
\begin{figure} \centering    
\subfigure[] {
 \label{velocity-Cluster-lifetime}     
\includegraphics[width=0.7\columnwidth]{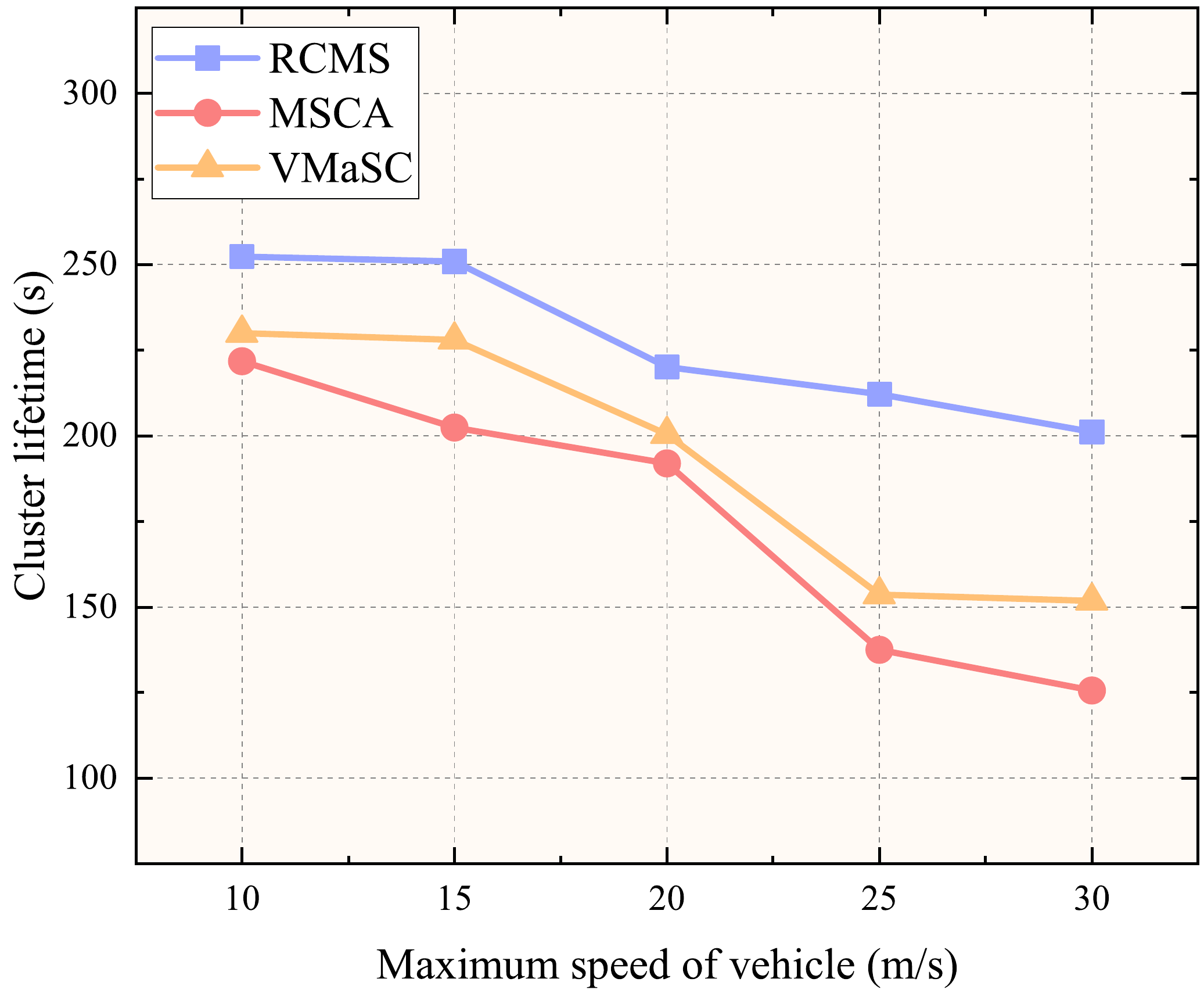}  
}     

\subfigure[] {
 \label{TTI-Cluster-lifetime}     
\includegraphics[width=0.7\columnwidth]{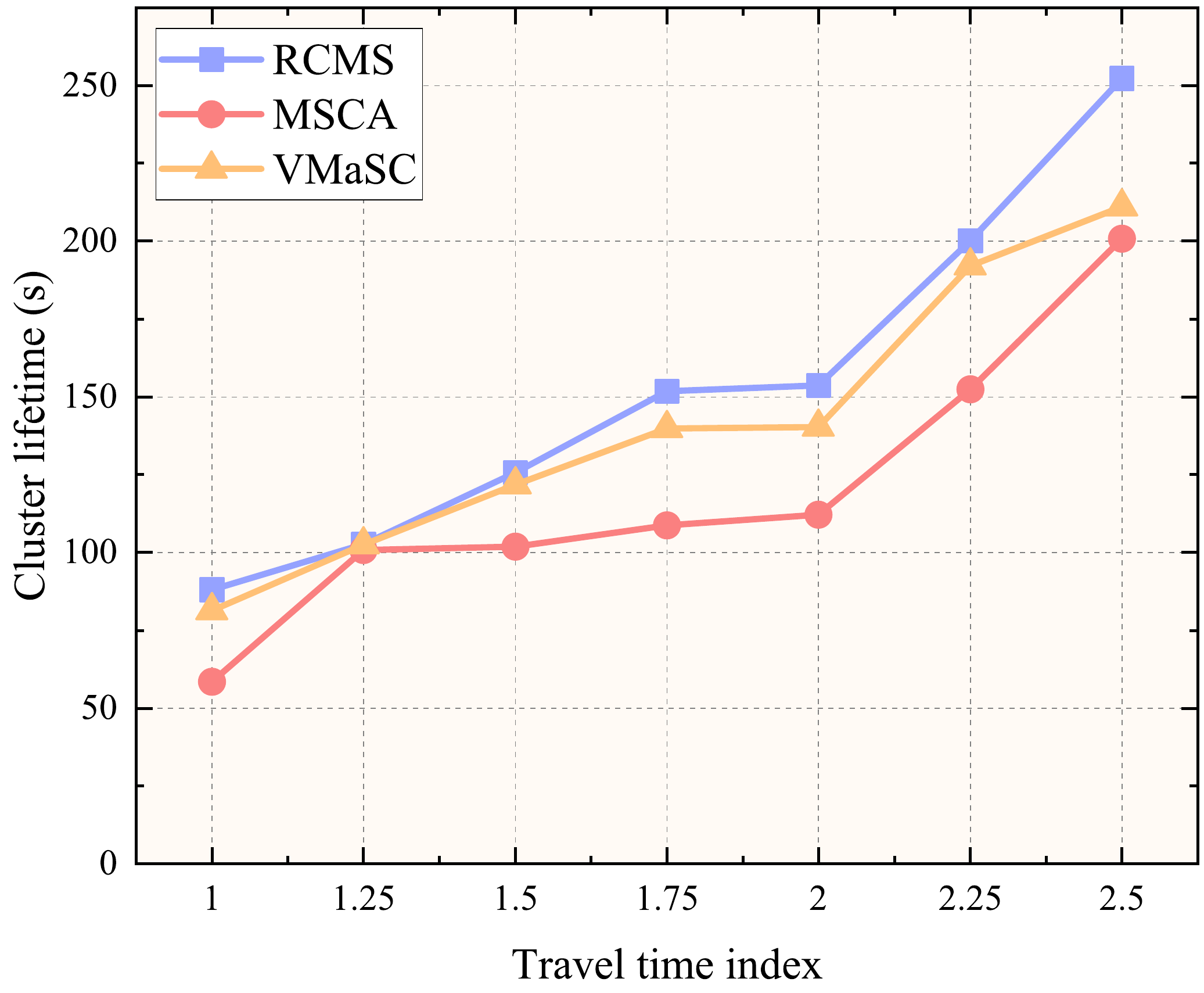}  
}     
\caption{RCMS performance of cluster lifetime. Relationship between (a) cluster lifetime and maximum speed of vehicle, (b) cluster lifetime and TTI}     
\label{fig:lifetime}     
\end{figure}
We propose a region-based collaborative management scheme for dynamic clustering, named RCMS. We compare RCMS with the two existing clustering algorithms, the vehicular multihop algorithm for stable clustering (VMaSC) \cite{ucar2015multihop} and the mobility-based stability-based clustering algorithm (MSCA) \cite{ren2016new}. VMaSC utilizes the relative mobility metric to estimate the average relative speed with respect to neighboring vehicles. Based on the vehicle's moving metrics,  MSCA estimates the lifetime of communication links. Hence, we adopt the two algorithms as our experiments' baselines.

As shown in Fig.~\ref{velocity-Cluster-lifetime}, under different maximum speeds, RCMS achieves the longest clustering lifetime, followed by the MSCA and VMaSC, respectively. Fig.~\ref{TTI-Cluster-lifetime} illustrates the clustering lifetime in different TTI. Obviously, compared with MSCA and VMaSC, RCMS performs better. The reason explained as follows. VMaSC relies on the current direction of vehicle movement to establish clusters, while MSCA considers the effect of relative position and link lifetime. On the other hand, RCMS adopts the state resemblance prediction of vehicles with the influence of the traffic environment and can effectively predict vehicle occurrence in the near future. Therefore, it is longer in terms of cluster lifetime.

\begin{figure} \centering
\subfigure[] { 
\label{Totaltime-reconstruction}     
\includegraphics[width=0.7\columnwidth]{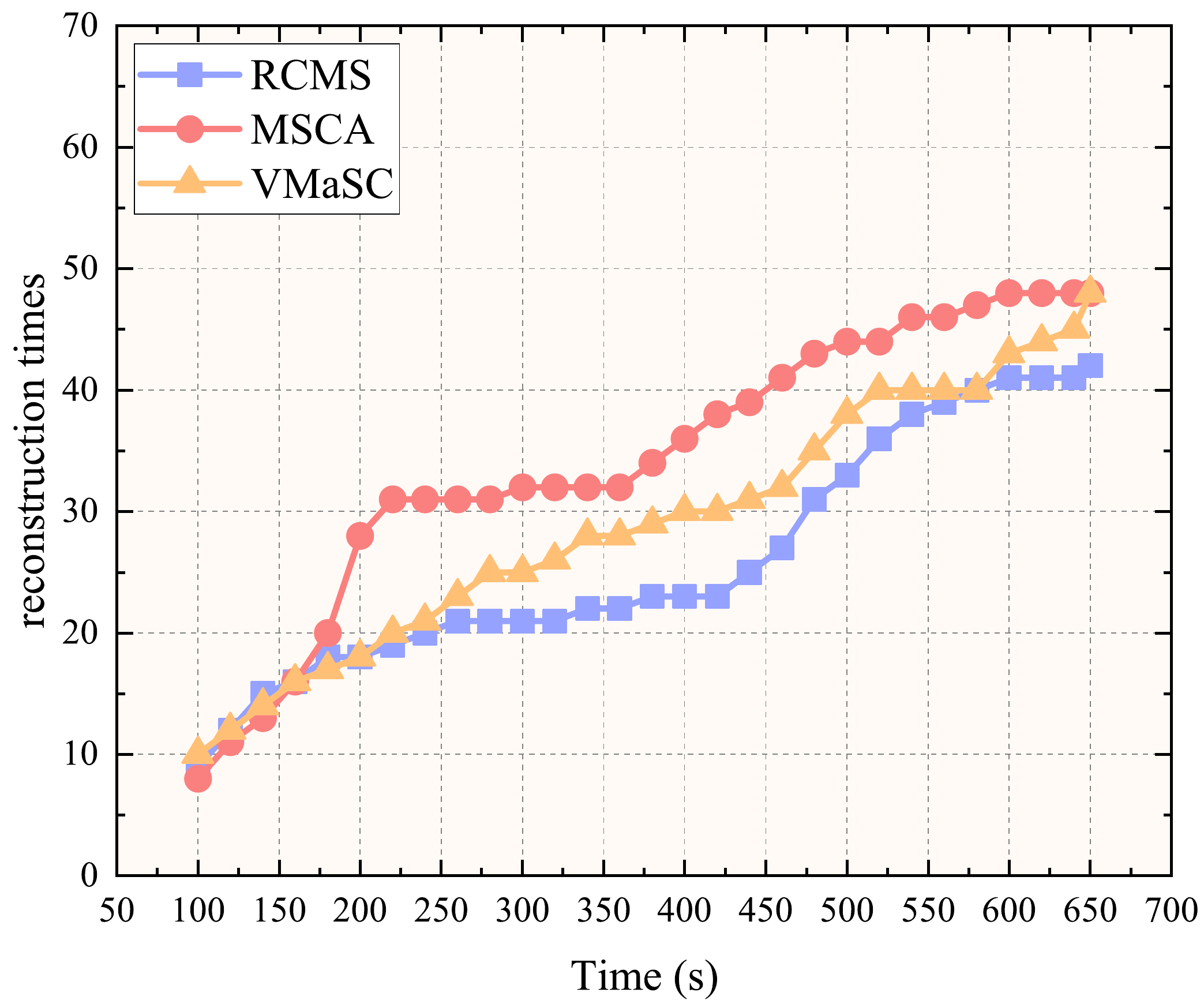}
}   

\subfigure[] { 
\label{TTI-reconstruction}     
\includegraphics[width=0.7\columnwidth]{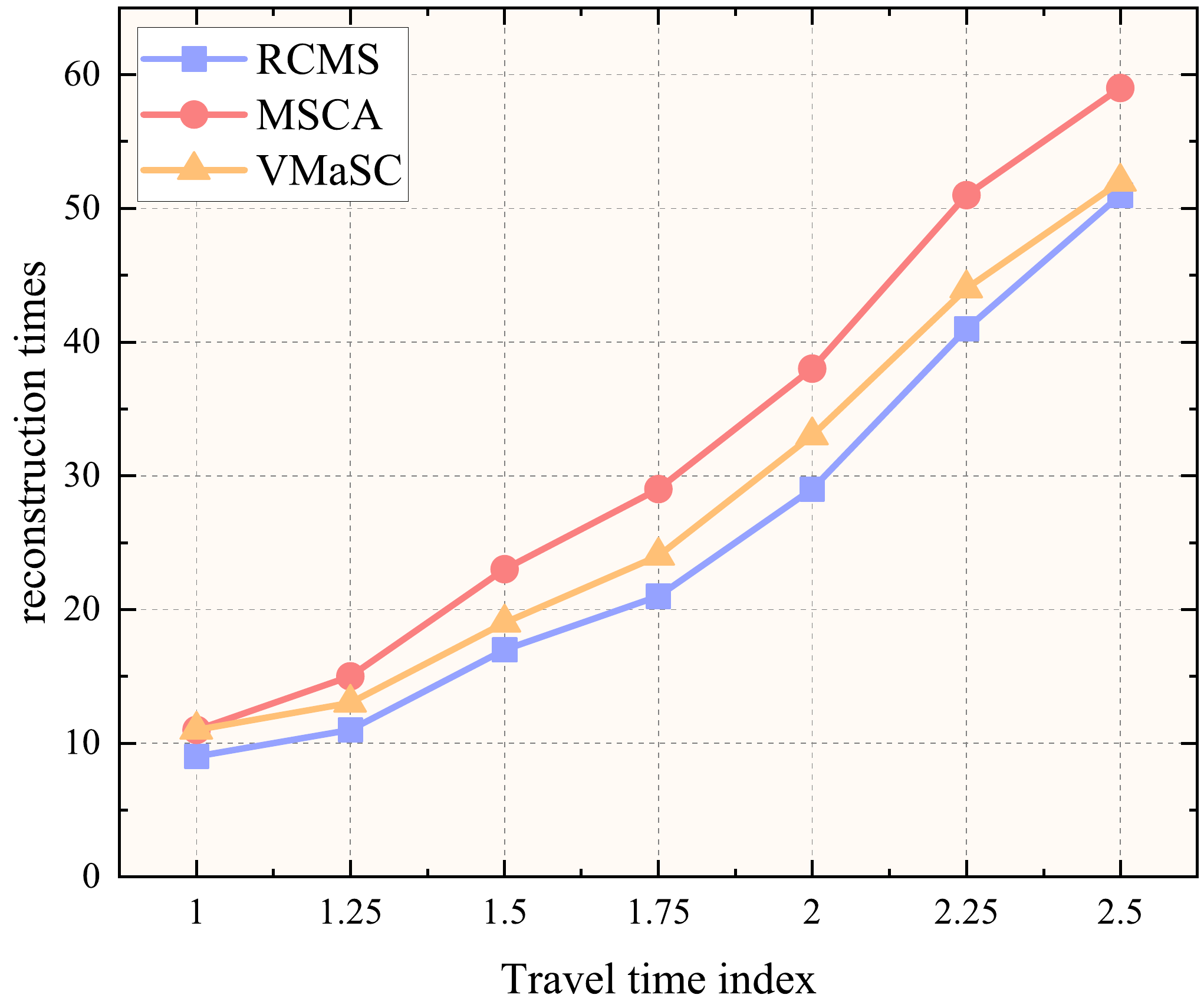}     
}    
\caption{RCMS performance of reconstruction times. Relationship between (a) reconstruction times and time, (b) reconstruction times and TTI}
\label{fig:reconstruction}
\end{figure}
Fig.~\ref{Totaltime-reconstruction} illustrates the number of cluster reconstructions at different moments. It is observed that the number of reconstructions increases with time. RCMS's reconstruction is lower than MSCA and VMaSC. Similarly, Fig.~\ref{TTI-reconstruction} reveals the relationship between the number of clustering reconstructions and TTI. The increase in the TTI directly leads to an increase in reconstruction times. Nonetheless, the reconstruction times of RCMS are less than those of MSCA and VMaSC.

\begin{figure}
\centering
\subfigure[] { 
\label{Totaltime-overlap}     
\includegraphics[width=0.71\columnwidth]{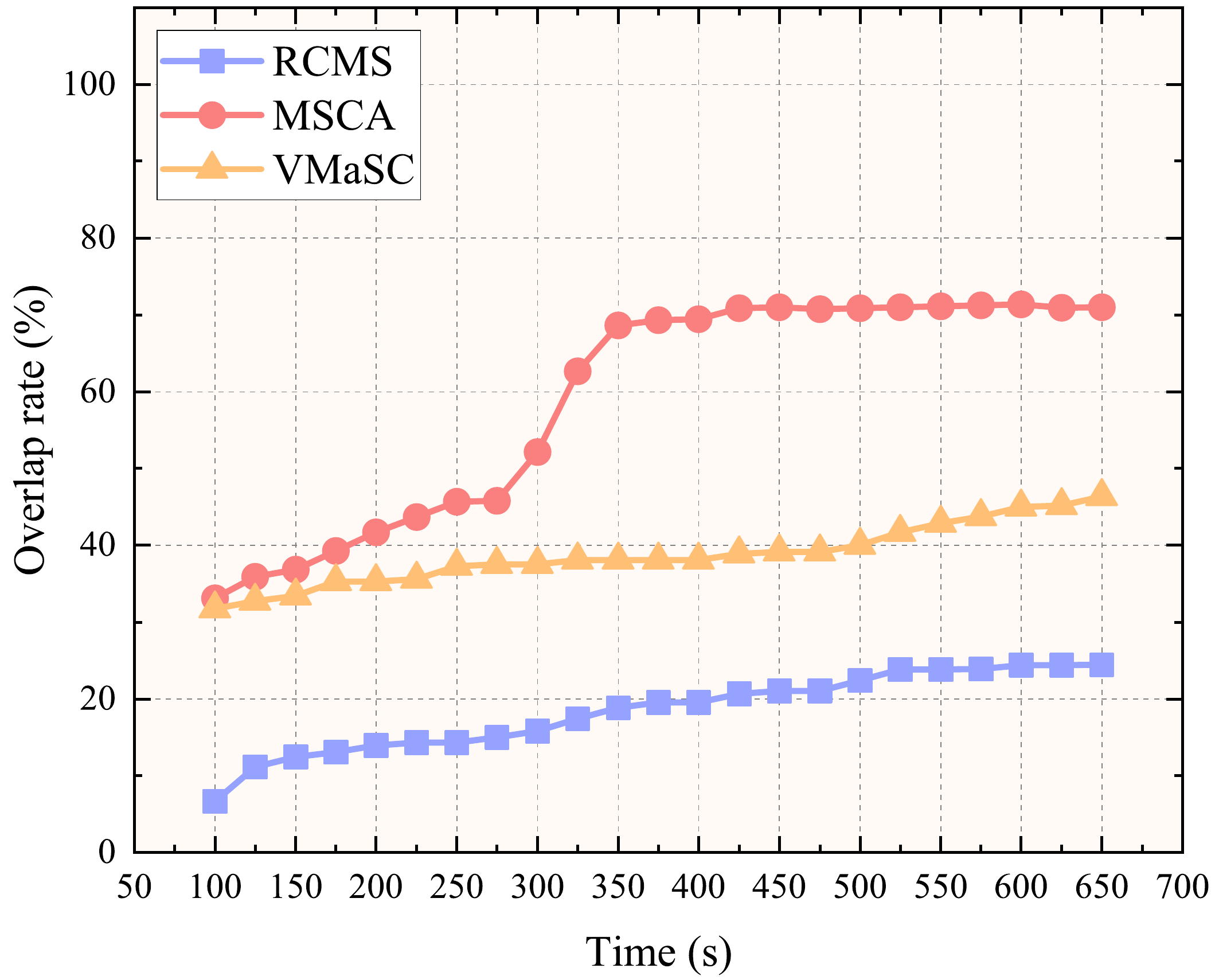}     
}   

\subfigure[] { 
\label{TTI-overlap}     
\includegraphics[width=0.71\columnwidth]{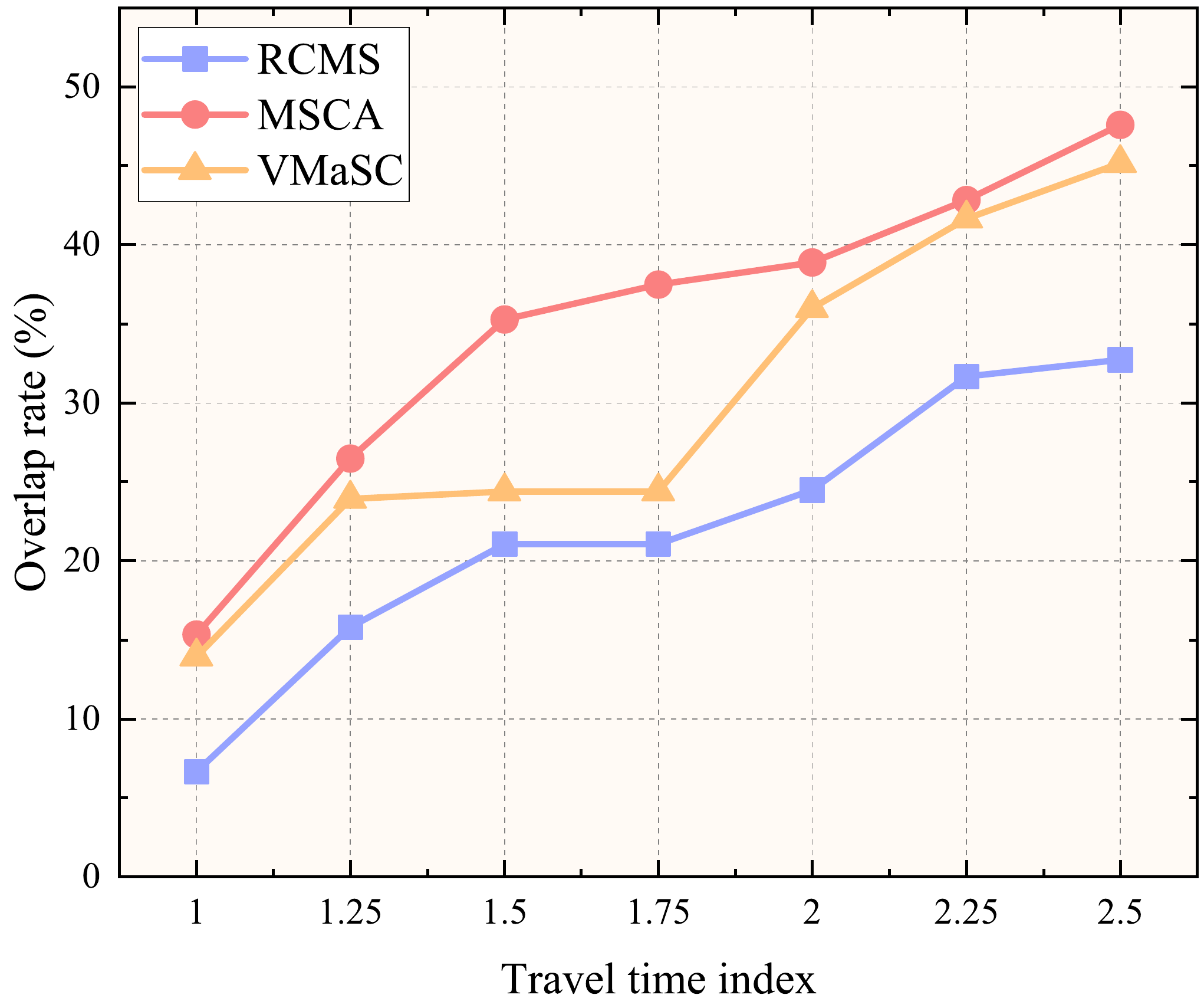}     
}   
\caption{RCMS performance of overlap. Relationship between (a) overlap rate and time, (b) overlap rate and TTI}
\label{fig:overlap}
\end{figure}
Fig.~\ref{Totaltime-overlap} shows that the overlap rate increases with time. The overlap rate of RCMS is between 9$\%$ and 26$\%$, much less than those of MSCA and VMaSC (between 35$\%$ and 61$\%$, 36$\%$ and 71$\%$, respectively). Fig.~\ref{TTI-overlap} illustrates the relationship between overlap rate and TTI. Similarly, RCMS's outperform both MSCA and VMaSC. The reason is that RCMS considers and proposes a joint control scheme of network connectivity and communication overlap. Hence, RCMS can achieve better connectivity with less overlap through fewer reconstructions.

\subsection{Communication Performance Evaluation}

In order to verify the connectivity performance of region-based collaborative management scheme for dynamic clustering, we use the region-based collaborative management scheme described in Section \Rmnum{5}. In this paper, we execute intra-cluster and inter-cluster operations to maintain the stability of the network structure. Based on the stable network structure, we can transmit a message to the designated destination through the relay nodes as the core vehicles and their corresponding gateway nodes. To precisely evaluate the communication performance, we use clustering-based routing protocol and non-clustering-based routing protocol as the baselines. For the clustering-based routing protocols, we choose the clustering-based directional routing protocol (CBDRP) \cite{song2010cluster}, in which the vehicle at the cluster center is selected as the cluster head, and member vehicle's routing messages need to be transmitted to the cluster head first. Then the cluster head determines the routing path of the message. For the non-clustering-based routing protocols, we select GPSR \cite{bala2015scenario}, which uses the store-carry-forward routing mode to transmit data packets to neighboring vehicles without constructing a routing path directly.

\begin{figure} \centering    
\subfigure[] {
 \label{TTI-DPDR}     
\includegraphics[width=0.7\columnwidth]{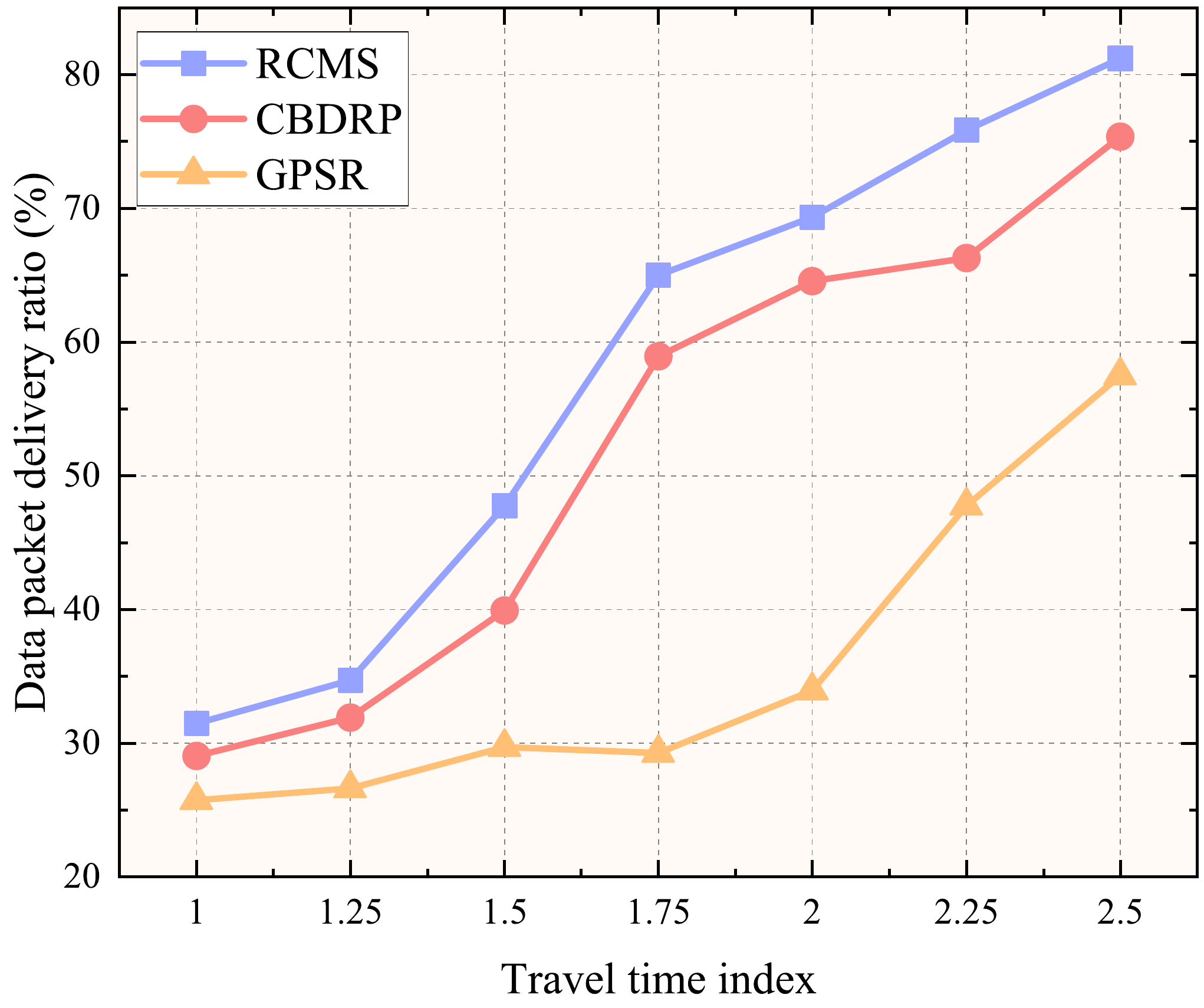}  
}     

\subfigure[] {
 \label{NDP-DPDR}     
\includegraphics[width=0.7\columnwidth]{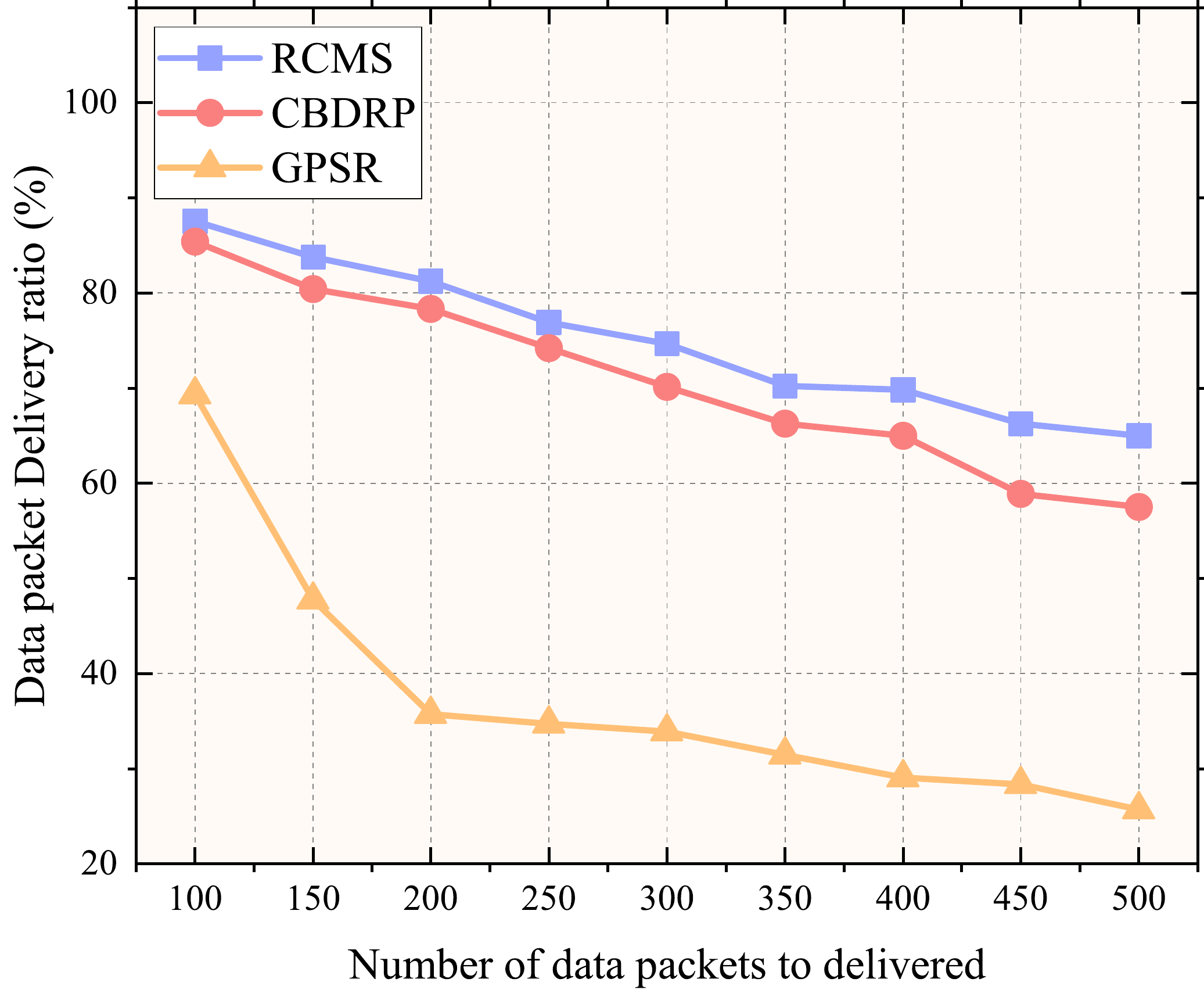}  
}  
\caption{Communication performance. Relationship between (a) DPDR and TTI. (b) DPDR and number of the packet to delivered} 
\label{fig:DPDR}     
\end{figure}
The relationship between data packet delivery rate (DPDR) and TTI is illustrated in Fig.~\ref{TTI-DPDR}. We can observe that as the TTI index increases, the vehicle density increases accordingly, and so does the data packet delivery rate. However, the growth slope before 1.75 of TTI is significantly greater than that after 1.75. As shown in the Fig.~\ref{NDP-DPDR}, when the number of data packets to be delivered (NDPD) increases, the DPDR decreases. Specifically, RCMS's DPDR is between 74$\%$ and 88$\%$, CBDRP's DPDR is between 85$\%$ and 58$\%$, and GPSR's DPDR is between 70$\%$ and 29$\%$. Therefore, RCMS outperforms CBDRP and GPSR.

\begin{figure} \centering
\subfigure[] { 
\label{TTI-IDPV}     
\includegraphics[width=0.7\columnwidth]{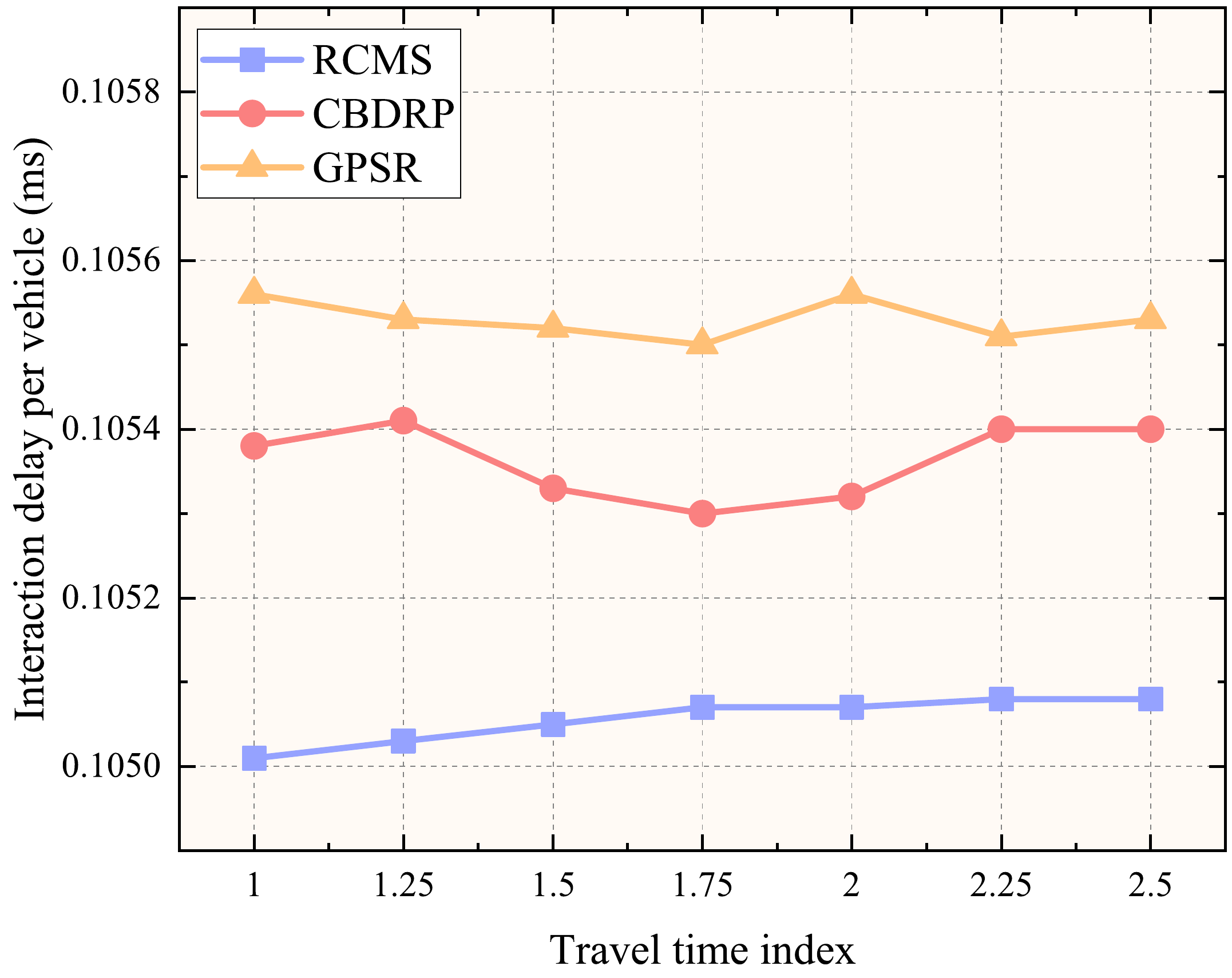}     
}  

\subfigure[] { 
\label{NDP-IDPV}     
\includegraphics[width=0.7\columnwidth]{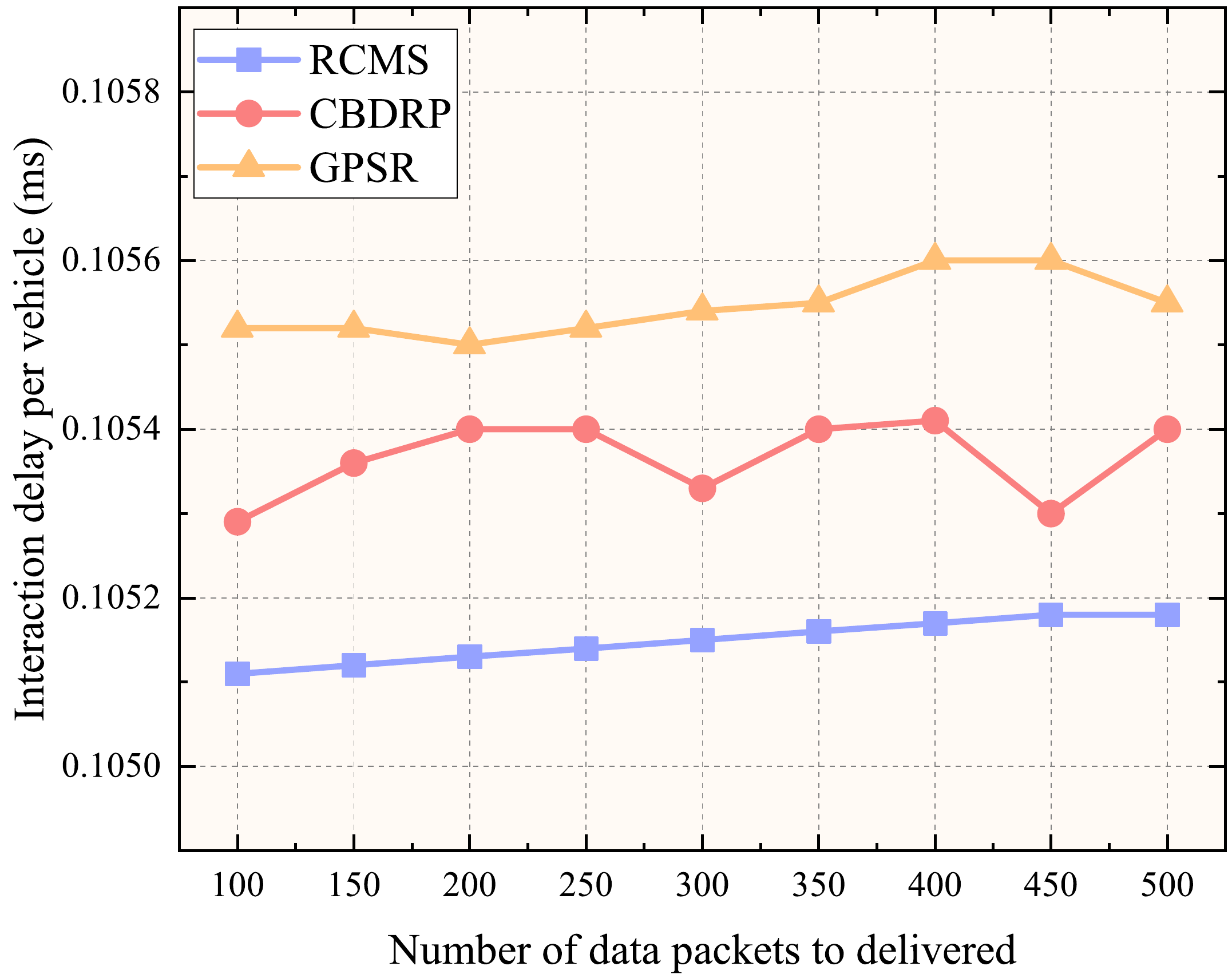}     
}   
\caption{Communication performance. Relationship between (a) interaction delay and TTI. (b) interaction delay and number of the packet to delivered. }
\label{fig:IDPV}
\end{figure}

Fig.~\ref{TTI-IDPV} demonstrates the relationship between the interaction delay and TTI. The interaction delay of RCMS is lower than those of CBDRP and GPSR. The GPSR routing algorithm is an algorithm that uses geographic location information to implement routing and a greedy algorithm to build a routing scheme, so its interaction delay is higher than CBDRP and RCMS. In contrast, RCMS adopts the distance information of vehicle nodes in bi-direction. Therefore, the interaction delay of RCMS is lower than that of CBDRP. Also, Fig.~\ref{NDP-IDPV} illustrates the relationship between interact delay and NDPD. The same result can be observed that the interaction delay of RCMS is lower than those of CBDRP and GPSR.



\section{CONCLUSION}

Green communications in VANETs have to take two important issues into consideration: transmission techniques and communication statues. To cope with such issues in this paper, we propose a region-based collaborative management scheme for dynamic clustering in VANETs. Specifically, for adapting to the traffic demands and communication requirements, we utilize a hierarchical network structure by grouping the vehicles into clusters. The vehicles within the same cluster have similar trajectory features determined by the state resemblance prediction (SRP) model. To maintain the stability of network structure, we propose the region-based collaborative management scheme (RCMS) to consider the relationship between network connectivity and communication overlap jointly. Based on the consideration, we provide intra-region and inter-region operations and aim to achieve better network quality and lower communication overhead. We conduct extensive experiments that validate the efficiency of the region-based collaborative management scheme for dynamic clustering.


%





\ifCLASSOPTIONcaptionsoff
  \newpage
\fi



%
\bibliographystyle{IEEEtran}
\bibliography{ref.bib}
\end{document}